\documentclass[12pt,showpacs,amsfonts]{revtex4-1}
\usepackage{amsmath}
\usepackage{latexsym}
\usepackage{float}
\usepackage{amssymb}
\usepackage{graphicx}
\usepackage{textcomp}
\usepackage{hyperref}

 2
\def\ba{\begin{eqnarray}}
\def\ea{\end{eqnarray}}
\def\be{\begin{equation}}
\def\ee{\end{equation}}

\def\bm{\begin{math}}
\def\me{\end{math}}

\newcommand{\dummy}

\begin{document}
\title{Dimensionality Dependence of Aging in Kinetics of Diffusive Phase Separation: Behavior of order-parameter 
autocorrelation}
%\vskip 0.5cm
\author{Jiarul Midya, Suman Majumder and Subir K. Das$^{*}$}
\affiliation{Theoretical Sciences Unit, Jawaharlal Nehru Centre for Advanced Scientific Research,
 Jakkur P.O., Bangalore 560064, India}

\date{\today}

\begin{abstract}
~Behavior of two-time autocorrelation during the phase separation in solid binary mixtures are studied via 
numerical solutions of the Cahn-Hilliard equation as well as Monte Carlo simulations of the Ising model. 
Results are analyzed via state-of-the-art methods, including the finite-size scaling technique. 
Full forms of the autocorrelation in space dimensions $2$ and $3$ are obtained empirically. The long time 
behavior are found to be power-law type, with exponents unexpectedly higher than the ones for the ferromagnetic 
ordering. Both Chan-Hilliard and Ising models provide results consistent with each other. 
\end{abstract} 


\maketitle
\section{Introduction}
~Properties of a nonequilibrium system change with growing age \cite{Zannetti}. 
Understanding of such aging phenomena is of fundamental importance in all 
branches of science and technology. There have been serious activities 
on this issue concerning living \cite{santos, costa} as well as nonliving matters, especially in problems related 
to domain growth \cite{Zannetti, DFisher, Mazenko, Liu, Desai, Mazenko_recnt, Barkema, puri, Gemmert, Shaista, Majumder,Jiarul}
and glassy dynamics \cite{glass1,glass2,glass3,glass4,glass5}. 
Among other quantities, aging phenomena is studied via the two-time autocorrelation function \cite{DFisher} 
\begin{eqnarray}\label{auto_Cr}
 C(t,t_w)=\langle \psi (\vec{r}, t)\psi(\vec{r}, t_w)\rangle-\langle\psi(\vec{r}, t)\rangle\langle\psi(\vec{r}, t_w)\rangle.
\end{eqnarray}
In Eq. (\ref{auto_Cr}), $\psi$ is a space ($\vec{r}$) and time dependent order parameter, 
$t_w$ is the waiting time or age of the system and $t$ ($>t_w$) is the observation time.
\par
~ In phase ordering systems \cite{Bray}, though time translation invariance is broken,
$C(t, t_w)$ is expected to exhibit scaling with respect to $t/t_w$. 
Important examples are ordering of spins in a ferromagnet, kinetics of phase separation in a
binary ($A+B$) mixture, etc., having been quenched to a temperature ($T$) below the critical value ($T_c$),
from a homogeneous configuration. Though full forms are unknown even for
very simple models, asymptotically $C(t,t_w)$ is expected to obey power-law scaling behavior as \cite{DFisher, Liu} 
\begin{eqnarray}\label{scle_Cr}
 C(t, t_w) \sim x^{-\lambda}; ~ x=\ell/\ell_w.
\end{eqnarray}
In Eq. (\ref{scle_Cr}), $\ell$ and $\ell_w$ are the average sizes of domains, formed by spins or particles of 
similar type, at times $t$ and $t_w$, respectively. Typically $\ell$ and $t$ are related to each other via power-laws. 
\par
~For nonconserved order-parameter dynamics, e.g., ordering in a ferromagnet, such scaling has been observed and 
the values of the exponent $\lambda$ have been accurately estimated \cite{Liu, Jiarul} 
in different space dimensions $d$. There the exponents follow the bounds 
\begin{eqnarray}\label{bound_FH}
\frac{d}{2} \leq \lambda \leq d,
\end{eqnarray}
predicted by Fisher and Huse (FH) \cite{DFisher}. In kinetics of phase separation in solid 
mixtures, for which the order parameter is a conserved quantity, the state of understanding is far from satisfactory, 
due to theoretical as well as computational difficulties. There exist reports \cite{Gemmert} of violation of scaling 
with respect to $t/t_w$. The latter observation, our results indicate, is due to the fact that scaling is achieved 
for $t_w\gg 1$. Access to such long time is constrained by inadequate computational resources. 
This difficulty, in some studies \cite{Majumder, Barkema}, might have led to the conclusion about incorrect 
values of $\lambda$.
\par
In an important work, Young et al. \cite{Desai} put a more general lower bound on $\lambda$, valid irrespective 
of the conservation of the total order parameter, as 
\begin{eqnarray}\label{YRD_bound}
\lambda \geq \frac {\beta+d}{2},
\end{eqnarray}
where $\beta$ is the exponent for small wave-vector power-law enhancement of equal time structure factor which, 
depending upon the dynamics, becomes important for $t_w\gg 1$, as stated below. In nonconserved dynamics, 
$\beta=0$ and so the lower bound in Eq. (\ref{bound_FH}) is recovered. For conserved order parameter dynamics, 
on the other hand, $\beta=4$ in both $d=2$ and $3$ at late time. Thus, the upper bound in Eq. 
(\ref{bound_FH}) is violated. Simulations of the Cahn-Hilliard (CH) equation \cite{Bray}
\begin{eqnarray}\label{ch_eq}
\frac{\partial \psi(\vec{r}, t)}{\partial t}=-\nabla^{2}\Big[\psi(\vec{r},t)+\nabla^{2}\psi(\vec{r},t)-\psi^{3}(\vec{r},t)\Big],
\end{eqnarray}
by Young et al. \cite{Desai}, observed $\lambda > 3$ in $d=2$, consistent with Eq. (\ref{YRD_bound}). From these 
simulations, the authors, however, did not accurately quantify $\lambda$; scaling of $C(t, t_w)$ with 
respect to $t/t_w$ was not demonstrated; focus was rather on the sensitivity of the 
aging dynamics to the correlations in the initial configurations. Situation is far 
worse in $d=3$, with respect to the CH equation as well as the Ising model 
\cite{Zannetti, Bray}
\begin{eqnarray}\label{ham_isi}
 H = -J\sum_{<ij>}S_i S_j;~S_{i}=\pm1;~J>0.
\end{eqnarray}
\par
~ In this paper, we study both CH equation and the Ising model, used for understanding diffusive phase separation as 
in solid mixtures, in $d=2$ (on regular square lattice) and $d=3$ (on simple 
cubic lattice), via extensive simulations, to quantify the decay of $C(t, t_w)$. 
We observe scaling of $C(t,t_w)$ with respect to $x$ in which the power-law of Eq. (\ref{scle_Cr}) is realized 
for large $x$. Via computations of the instantaneous exponent 
\cite{Huse, Amar, Suman_Ising}
\begin{eqnarray}\label{ins_lam}
\lambda_{i}= -\frac{d \ln [C(t, t_w)]}{d \ln x},
\end{eqnarray}
and application of the finite-size scaling technique \cite{MFisher,MCbook}, we find that $\lambda \simeq 3.6$ 
in $d=2$ and $\simeq 7.5$ in $d=3$. Though these numbers respect the bounds in Eq. (\ref{YRD_bound}), the high 
value in $d=3$ is surprising. But this comes from both CH and Ising models, from various reliable analyses. 
Furthermore, a general analytical form for the full scaling functions has been obtained empirically.

\section{Methods}
\par
~One numerically solves the CH equations on a regular lattice, usually via Euler discretization method. With the Ising
model, phase separation kinetics in a solid binary mixture is studied via the Kawasaki exchange 
Monte Carlo (MC) \cite{MCbook} simulations, to be referred to as KIM. An up spin ($S_i=+1$), for this problem, 
may correspond to an $A$ particle and a down spin ($-1$) to a $B$ particle. In this MC scheme, one randomly chooses 
a pair of nearest neighbor spins and tries their position exchange. The moves are accepted according to 
standard Metropolis algorithm \cite{MCbook}. Due to the coarse-grained nature of the CH equation, as opposed 
to the atomistic Ising model, one can explore large effective length in simulations. The order parameter 
in Eq. (\ref{ch_eq}) corresponds to a coarse-graining \cite{Onuki} of the Ising spins, typically over 
the equilibrium correlation length $\xi$. Then, a positive value of $\psi$ means an $A$-rich region and 
for a $B$-rich region, $\psi$ will have a negative number. For the calculation of $C(t, t_w)$, we have used binary numbers 
$+1$ and $-1$, for both the models.
\par
~ The average domain length, $\ell$, in our simulations was measured from the first moment of domain size distribution, 
$P(\ell_d, t)$, as \cite{Suman_Ising}
\begin{eqnarray}\label{leng_form}
\ell=\int \ell_d~P(\ell_d, t)~d\ell_d,
\end{eqnarray}
where $\ell_d$ is the distance between two successive domain boundaries in any direction. Throughout the paper, all 
lengths are presented in units of the lattice constant $a$. 
In MC simulations, time is counted in units of Monte Carlo steps (MCS), each MCS consisting of $L^d$ trial moves, 
where $L$ is the linear dimension of a periodic square or cubic system. 
In CH equation, $t$ is expressed dimensionless units \cite{Binder_puri}. All results are presented after averaging over 
at least $50$ initial configurations, for quenches from random initial configurations to $T=0.6T_c$.
\section{Results}
\begin{figure}[htb]
\centering
\includegraphics*[width=0.35\textwidth]{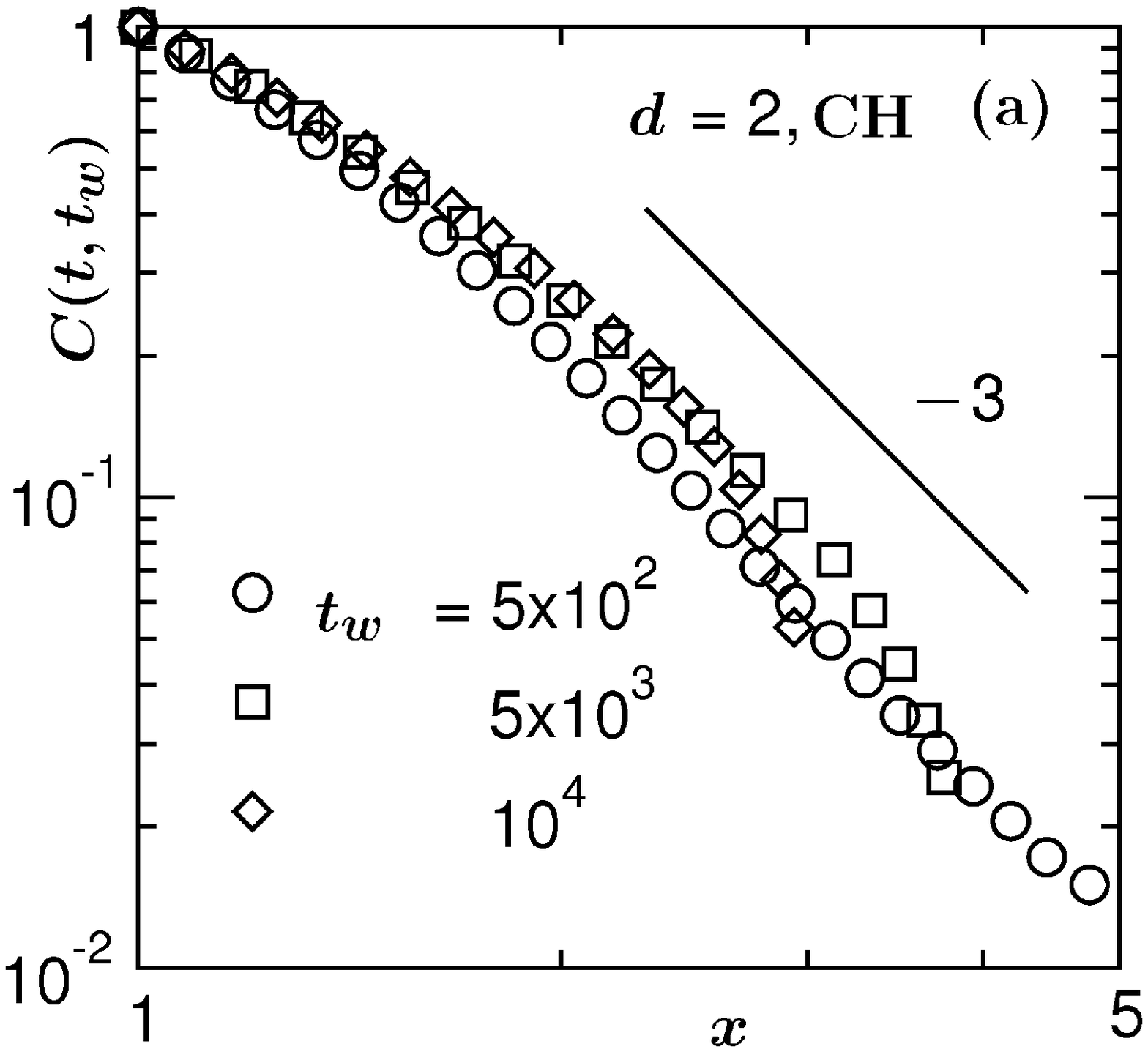}
\vskip 0.50 cm
\includegraphics*[width=0.35\textwidth]{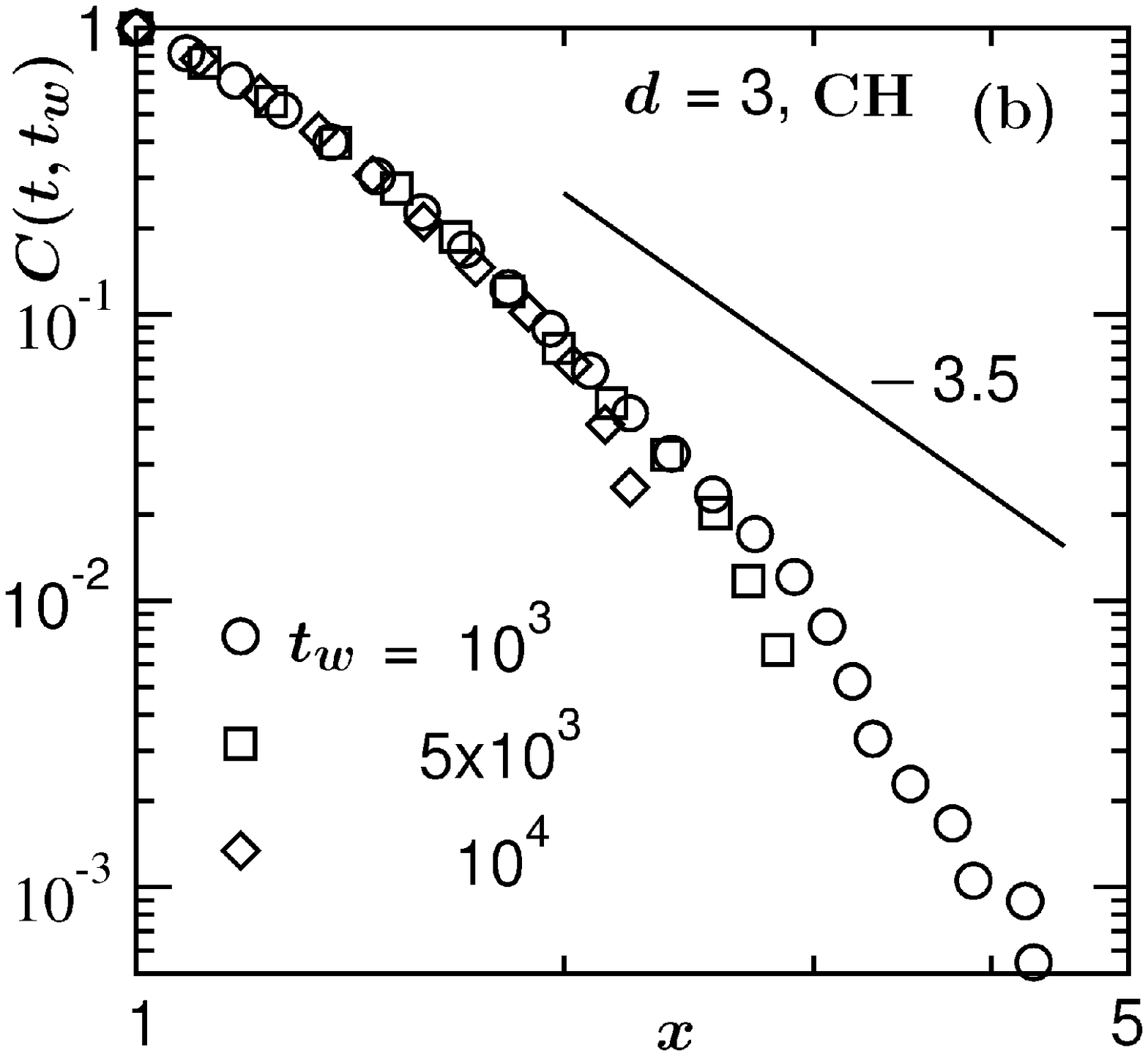}
\caption{\label{fig1} (a) Autocorrelation function, $C(t, t_w)$, from the  $d=2$ Cahn-Hilliard model, are plotted vs 
$x$ ($ = \ell/\ell_w$), for different values of $t_w$. The solid line there corresponds to a power-law decay with exponent 
$3$. (b) Same as $(a)$ but for the $d=3$ CH model. The solid line there has a power-law decay exponent $3.5$.
The system sizes used are $L=256$ ($d=2$) and $200$ ($d=3$).
}
\end{figure}
\par
~In Fig. \ref{fig1}(a), we present the plots of $C(t, t_w)$, vs $x$, for different values of $t_w$, 
from the solutions of CH model in $d=2$. 
As seen, one needs large enough value of $t_w$ to observe appropriate scaling behavior,
compared to ordering in ferromagnets \cite{Jiarul}. Between the two data sets with largest values of $t_w$, 
the deviation from each other, for large $x$, is due to the finite-size effects. 
Similar plots for the $d=3$ CH model are presented in Fig. \ref{fig1}(b). Here all $t_w$s are large enough, 
providing good scaling. Again, deviation from the master curve, starting at different values of 
$x$ for different $t_w$, are primarily related to the finite-size effects. In both these figures, 
\ref{fig1}(a) and \ref{fig1}(b), the system sizes are kept fixed, only the values 
of $t_w$ are varied. A similar observation, with respect to the above mentioned deviation for 
different choices of $t_w$, can be made, when, for same value of $t_w$, data are presented for different system sizes. 
\par
~In the scaling parts, both in Fig. \ref{fig1}(a) and Fig. \ref{fig1}(b), continuous bending is observable, in these log-log 
plots. Thus, power-laws, if exist, carry corrections. The solid lines in these figures are power-law decays 
with exponents $3$ and $3.5$, respectively, corresponding to the bounds in Eq. (\ref{YRD_bound}). 
For large $x$, simulation data in $d=2$ appear reasonably consistent 
with the bound. The asymptotic exponent, in $d=3$, on the other hand, 
appear much higher than $3.5$.
\begin{figure}[htb]
\centering
\includegraphics*[width=0.35\textwidth]{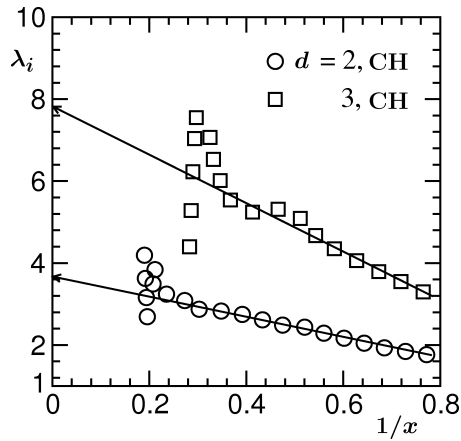}
\caption{\label{fig2} Instantaneous exponents $\lambda_i$ are plotted vs $1/x$. Results are shown only from the solutions
of the CH equations, in both $d=2$ and $3$. The solid lines are guides to the eye.
The $d=2$ data are for $t_w=5\times 10^3$ with $L=400$. In $d=3$ the numbers are $10^3$ and $200$.
}
\end{figure}
\par
~With the expectation that power laws indeed exist, in Fig. \ref{fig2} we present plots of instantaneous exponents 
\cite{Liu,Jiarul,Huse, Amar, Suman_Ising} $\lambda_i$, for both $d=2$ and $3$, vs $1/x$. 
In addition to providing $\lambda$, from the extrapolations to $x=\infty$, such exercise may be 
useful for obtaining crucial information on the full forms of $C(t,t_w)$. For $d=2$, the data are obtained for 
$t_w=5\times 10^3,$ and for $d=3$, the data correspond to $t_w=10^3$. 
In both the cases, the results appear reasonably linear \cite{Liu, Jiarul}. 
The solid lines there are extrapolations to $x=\infty$, accepting the linear trends. 
These indicate $\lambda \simeq 3.60$ in $d=2$ and $\simeq 7.80$ in $d=3$. 
Again, while the value in $d=2$ is consistent and close to the bound of Yeung \cite{Desai} et al., the observation of 
surprisingly high number in $d=3$ is certainly interesting. We intend to obtain more accurate values via appropriate 
finite-size scaling analyses \cite{MCbook, MFisher}. This is considering the fact that the choice of the regions 
in Fig.~\ref{fig2}, for performing least-square fitting, is not unambiguous due to finite-size effects and strong 
statistical fluctuations at large $x$. Also, for very small $x$ (data excluded), there is rapid decay of $C(t, t_w)$ 
related to the fast equilibration of domain magnetization. 
\par
~Since the corrections to the asymptotic decay laws are seen to be strong for finite $x$, a reasonable idea about the 
full forms of the decays is essential for accurate finite-size scaling analyses. Those, however, are nonexistent in the 
literature. Here we obtain the forms empirically.
\begin{figure}[htb]
\centering
\includegraphics*[width=0.35\textwidth]{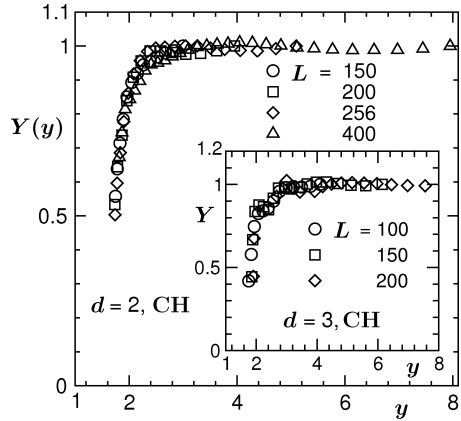}
\caption{\label{fig3} Finite-size scaling plot of $C(t,t_w)$ from $d=2$ CH model. The scaling function $Y$ is plotted 
vs $y$, using data from different system sizes. The optimum collapse of data, the presented one, was obtained for 
$\lambda=3.47$. The inset presents the same exercise for the CH model in $d=3$. Here the value of $\lambda$ 
is $7.30$.
See text for values of $t_w$.
}
\end{figure}
Assuming power-law behavior of the data sets in Fig. \ref{fig2}, we write
\begin{eqnarray}\label{ins_linear}
\lambda_i=\lambda-\frac{A_c}{x^{\gamma}}, 
\end{eqnarray}
where $A_c$ and $\gamma$ are constants. Combining Eq. (\ref{ins_linear}) with Eq. (\ref{ins_lam}), we obtain 
\begin{eqnarray}\label{emp_cr}
C(t, t_w) = C_0 \exp \Big(-\frac{A_c}{\gamma x^{\gamma}}\Big) x^{-\lambda}, 
\end{eqnarray}
$C_0$ being a constant. For finite-size scaling analysis, one needs to introduce a scaling function
\begin{eqnarray}\label{scale_fun}
 Y(y)=C(t, t_w) \exp\Big(\frac{A_c}{\gamma x^\gamma}\Big)x^{\lambda}; ~~y=L/\ell.
\end{eqnarray}
For appropriate choices of $A_c$, $\gamma$ and $\lambda$, one should obtain a master curve for $Y$, 
when data from different system sizes are used. The behavior of $Y$ should be flat in the finite-size unaffected region and 
a deviation from it will mark the onset of finite-size effects. 
\par
\begin{figure}[htb]
\centering
\includegraphics*[width=0.35\textwidth]{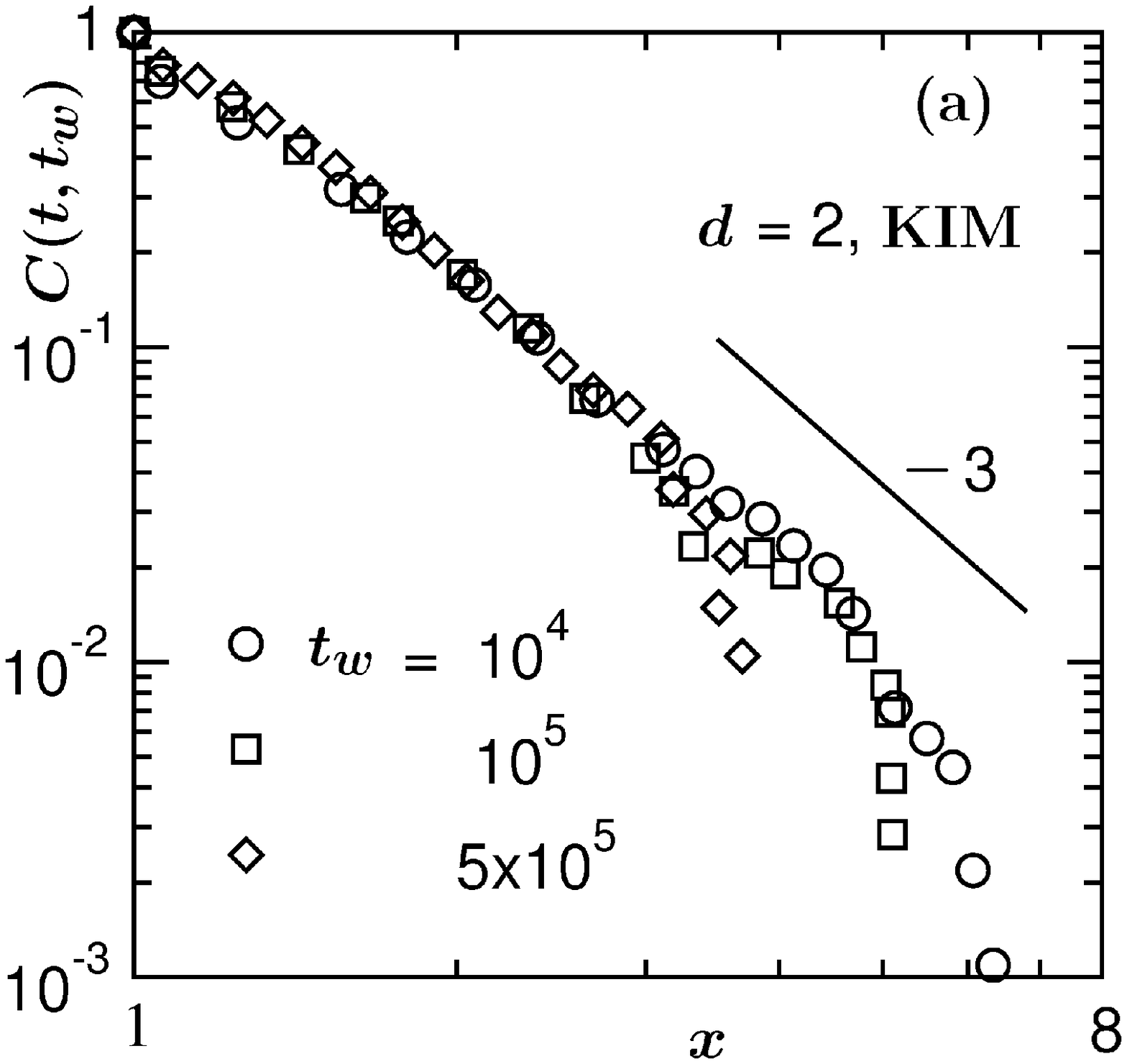}
\vskip 0.50 cm
\includegraphics*[width=0.35\textwidth]{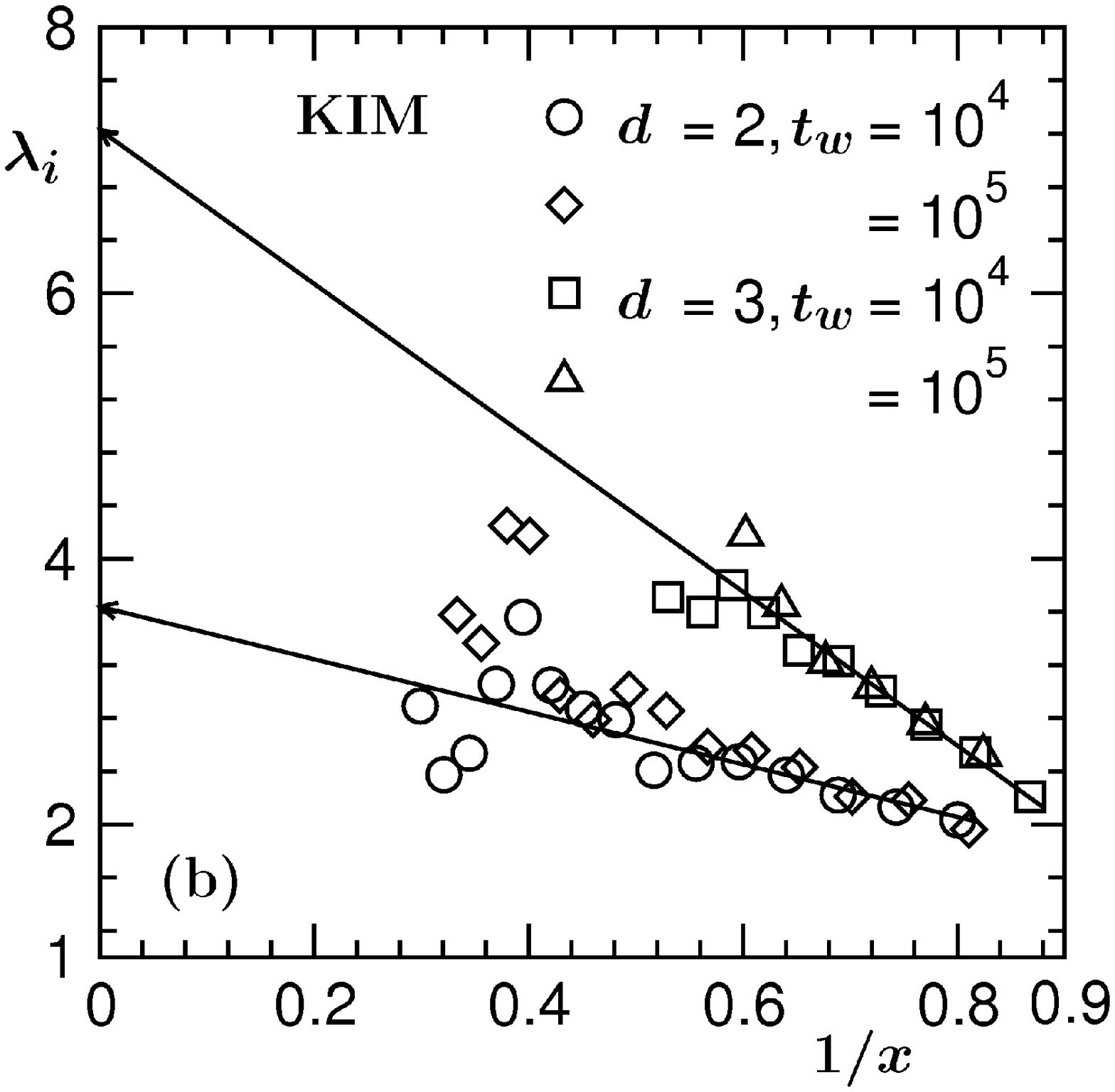}
\caption{\label{fig4} (a) Same as Fig. \ref{fig1}(a) but for the $d=2$ Ising model. (b) Same as Fig. \ref{fig2} but 
for Ising model. The oscillatory behavior of large $x$ data in $d=2$ is statistical fluctuation.
In $d=2$ the results are from $L=512$ and in $d=3$, we have used $L=100$.
}
\end{figure}

~ By examining the data in Fig. \ref{fig2} (also see Fig.~\ref{fig4}(b) for KIM), we fix $\gamma$ to $1$. 
In the main frame of Fig. \ref{fig3}, we show a finite-size scaling plot for data from the $d=2$ CH model. 
The presented results correspond to best collapse, obtained for $A_c= 2.25$ and $\lambda = 3.47$. The value of 
$t_w$ used for all the data sets is $10^4$. A similar exercise for the $d=3$ CH data is presented in the inset of 
Fig. \ref{fig3}. Again the data collapse looks quite reasonable and was obtained for 
$A_c=5.1$ and $\lambda = 7.30$. The value of $t_w$, in this case, was set to $10^3$. The reason behind choosing 
smaller value of $t_w$ in $d=3$, than in $d=2$, is computational difficulty. It is extremely difficult to 
accumulate data for further decades in time, starting from very high value of $t_w$, particularly in $d=3$. 
Nevertheless, this chosen value of $t_w$ falls within the scaling regime. Note that for similar temperatures, 
amplitude of growth is larger in $d=3$ and the scaling of $C(t, t_w)$ is related more closely to the value of $\ell_w$. 
In this connection we mention that the longest run lengths (associated with largest systems) for the CH model are 
$t=2\times10^6$ and $2\times10^5$ in $d=2$ and $3$, respectively; for the Ising model these numbers are 
$5\times10^7$ and $4\times10^6$.
\par
~ We now move to present results from KIM. In Fig. \ref{fig4}(a) we show the autocorrelations from different 
values of $t_w$ in $d=2$, for $L=512$. Scaling is poor for $t_w$ below $10^4$ MCS and so those results 
are excluded. Despite strong statistical fluctuations, it is recognizable that the decay of $C(t,t_w)$ 
in the latter part is on the higher side of the bound in Eq. (\ref{YRD_bound}), represented by 
the solid line. 
\begin{figure}[htb]
\centering
\includegraphics*[width=0.36\textwidth]{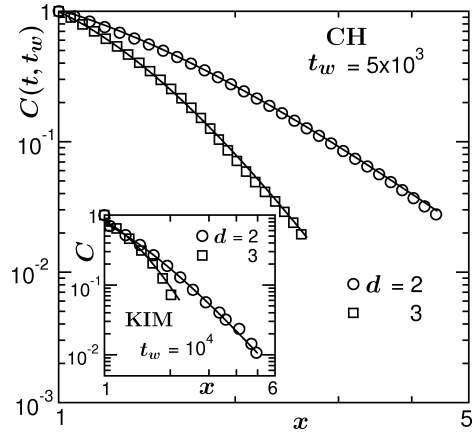}
\caption{\label{fig5} $C(t,t_w)$ are plotted vs $x$, for the CH model in $d=2$ and $3$. 
Inset shows corresponding results for the KIM. The solid lines are fits to the form in Eq. (\ref{emp_cr}),
with $\gamma=1$. The $t_w$ values are mentioned on the figure. We have discarded data
suffering from finite-size effects. The system sizes are
$L=400$ and $512$ for CH and Ising models in $d=2$, whereas, $L=200$ and $100$ in $d=3$.
}
\end{figure}
\par
~ In Fig. \ref{fig4}(b) we show the instantaneous exponents for the Ising model in $d=2$ and $3$, vs 
$1/x$. In each dimension, we have included two values of $t_w$ from the scaling regime. While results 
for different $t_w$s, in a particular dimension, are consistent with each other, finite-size effects appear earlier 
for larger value of $t_w$, as expected. Thus, for extrapolations to $x=\infty$, data sets with smaller $t_w$ 
are used. This exercise provides 
$\lambda \simeq 3.60$ and $\simeq 7.30$ in $d=2$ and $3$, respectively. These values are in agreement  
with the ones obtained for the CH model via various methods of analysis.
\par
~Finally, in Fig.~\ref{fig5} we show the fits of the simulation data to the form in Eq. (\ref{emp_cr}). 
The main frame is for the CH model in $d=2$ and $3$, whereas the inset contains
similar results from the KIM. The fits look quite satisfactory. This exercise provides 
$\lambda = 3.55$ and $3.76$ in $d=2$ for CH and Ising models, respectively. The numbers in 
$d=3$ are $7.64$ and $7.37$.
\section{Conclusions}
\par
~In conclusion, we have studied aging dynamics for the phase separation in solid binary mixtures via Cahn-Hilliard and 
Ising models. Results for the two-time autocorrelation, $C(t, t_w)$, are presented from simulations in both $d=2$ and $3$. 
Decays of $C(t,t_w)$ appear power law in large $x$ limit. The exponents for these power laws were obtained via 
various different analyses, including finite-size scaling, which is new for this purpose. For the finite-size 
scaling analysis, full forms of the autocorrelations were essential which we obtained empirically. All these 
methods provide consistent values of the decay exponent $\lambda$ for different models. 
These are $\lambda \simeq 3.6$ in $d=2$ and $\lambda \simeq 7.5$ in $d=3$, within $5\%$ error.
\par
~Very high value of $\lambda$ in $d=3$, far above the lower 
bound in Eq. (\ref{YRD_bound}), can be due to the fact that domains are more mobile in this dimension 
than in $d=2$. From the numbers obtained in $d=2$ and $3$, it may be tempting to predict a dimensionality 
dependence as $\lambda=f(d-1)$ with $f\simeq 3.75$. However, we caution the reader not to jump into such conclusion. 
Even though the influence of the dimension $d$ appear more important than in the bound of Eq. (\ref{YRD_bound}), 
the contribution of $\beta$ is significant, particularly at lower dimension. Results from other dimensions are 
necessary to make such a conclusion. In $d=1$ one should exercise the caution that $\beta$ ($=2$) has a different 
value\cite{SN_Majumder}.
\\
\vskip 0.1cm
\section{Acknowledgment}
%\textbf{Acknowledgment:}
\par 
~The authors acknowledge financial support from Department of Science and Technology, India, 
via Grant No SR/S2/RJN-13/2009. JM is grateful to the University Grants Commission for research fellowship and 
SM acknowledges Jawaharlal Nehru Centre for Advanced Scientific Research for financial support.
\vskip 0.2cm
${*}$ das@jncasr.ac.in

\end{document}